# Design of Reversible Counter

Md. Selim Al Mamun
Dept. of Computer Science and Engineering
Jatiya Kabi Kazi Nazrul Islam University
Trishal, Mymensingh-2220, Bangladesh

B. K. Karmaker
Dept.of Electronics and Communication Engineering
Jatiya Kabi Kazi Nazrul Islam University
Trishal, Mymensingh-2220, Bangladesh

*Abstract*—This article presents a research work on the design and synthesis of sequential circuits and flip-flops that are available in digital arena; and describes a new synthesis design of reversible counter that is optimized in terms of quantum cost, delay and garbage outputs compared to the existing designs. We proposed a new model of reversible T flip-flop in designing reversible counter.

## I. INTRODUCTION

R. Landauer [1] states that traditional logic operations dissipate heat due to the loss of information bits. It is proved that each bit of information loss generates kTln2 joules of heat energy; where k is Boltzmann's constant and T is the absolute temperature at which computation is performed. C. H. Bennett [2] showed that energy dissipation problem can be avoided if all the gates in the circuits are reversible. This is because reversible logic makes every step of computation to be completely reversible, so that no information is lost at any step of computation.

Research is going on reversible logic and a good amount of research work has been carried out in the area of reversible combinational logic. However, there is not much work in the area of sequential circuit like flip-flops and counters. A counter is a sequential circuit capable of counting the number of clock pulses that have arrived at its clock input. This paper proposes a novel of n bit reversible counter. The efficiency of the proposed design is proved with the help of proper theorems and algorithms.

The rest of the paper is organized as follows: Section 2 presents background on reversible logic. Section 3 describes related works on reversible counter. Section 4 describes the logic synthesis of our proposed reversible counter design and comparisons with other research works. Finally this paper is concluded with the Section 5.

## II. BACKGROUND ON REVERSIBLE LOGIC

This section focuses on the cost metrics used in this paper and describes some popular reversible gates along with their quantum representations.

### A. Cost Metrics

A reversible circuit can be synthesized in several ways, resulting different cost. This section outlines four cost metrics which are generally used to evaluate and compare reversible circuits.

*1) Gate Count:* This refers to the number of gates required to implement the circuit. This is used as a major cost metric in the evaluation of reversible sequential circuit [3]. But gate count is not a good metric for comparison as reversible gates are of different type and have different quantum costs [4].

*2) Garbage Output:* Some outputs are used only to maintain the reversibility of the circuit, but not result the final outputs nor are they used as input to other circuits. These unused outputs are known as garbage outputs.

*3) Delay:* The maximum number of gates in a path from any input line to any output line is considered as the delay of the circuit [5]. This paper used the logical depth as the measure of delay for reversible circuit proposed by Mohammadi and Eshghi [6].

*4) Quantum Cost:* The quantum cost of a reversible gate is the number of quantum gates or 1x1 and 2x2 reversible gates required to present the gate. The quantum costs of all reversible 1x1 and 2x2 gates are taken as unity [7].

This paper uses quantum cost, delay and the number of garbage bits as the cost metrics while comparing the proposed design with the existing results.

### B. Quantum Analysis of Popular Reversible Gates

Several reversible logic gates have been designed till now. Some popular reversible gates and their quantum equivalent diagrams are shown in Fig.1. Feynman gate (FG) [8] is the only 2*2 gate which has quantum cost 1. Among 3*3 gate Peres gate (PG) [9] has quantum cost 4, Selim Al Mamun (SAM) [10] gate has quantum cost 4 and Toffoli gate (TG) [11] has quantum cost 5.

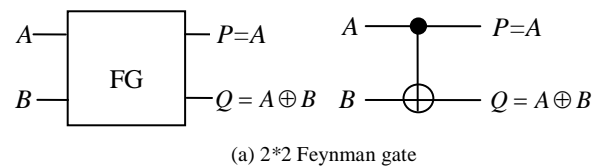

(a) 2*2 Feynman gate

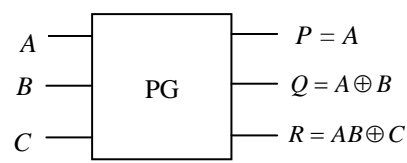

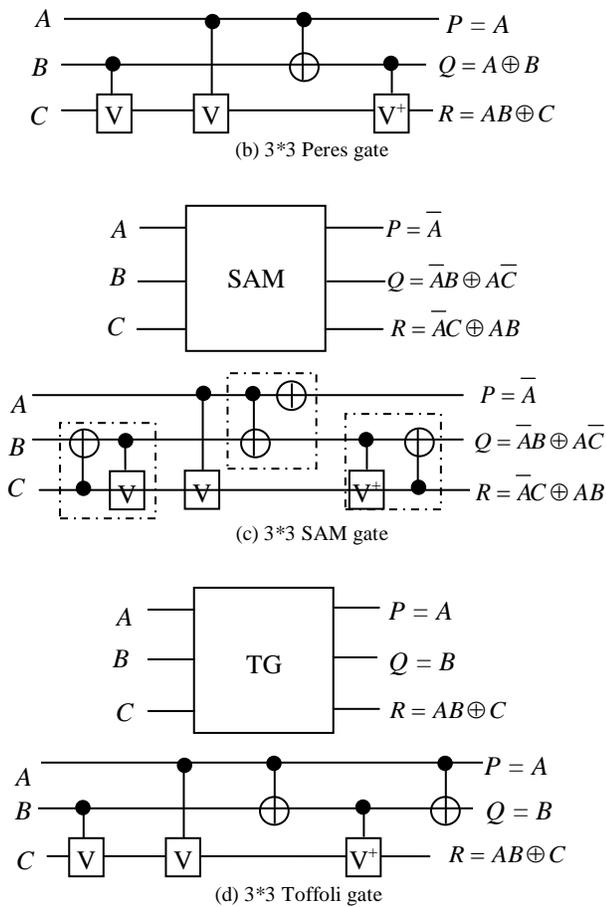

Fig. 1. Quantum analysis of popular reversible gates

### III. RELATED WORKS ON REVERSIBLE COUNTER

Researchers have worked on many ways on sequential circuit and work is still going on. This section reviews some previous implementation and sequential circuit designs. Researchers [10, 12-15] proposed the implementation of all types of latches, flip-flops and their master-slave design. These works opens a door to the implementation of large sequential circuit like counter.

The authors [16-17] carry the above success to the of design reversible counter. Some of the works are implemented by the replacement of latches and gates by their reversible counter parts. Recently Khan [16] used Positive Polarity Reed Muller (PPRM) expression to design synchronous counter. All these works suggest that there is scope for the design and implementation of large sequential circuits like counter.

### IV. DESIGN AND SYNTHESIS OF REVERSIBLE COUNTER

This section describes our proposed design for n bit counter. Designs for both the asynchronous counter andthe synchronous counter are presented here. While designing counter, this paper also proposed the design of reversible T flip-flop which is the building block of the counter. This paper presents the design of T flip-flop, gated T flip-flop and master slave T flip-flop.

### A. Proposed T Flip-flop

The characteristic equation of a T flip-flop is $T\bar{Q} \oplus Q\bar{T} = T \oplus Q$. A T Flip-flop can be realized by a single Feynman gate. Our proposed T flip-flop is shown in Fig.2.

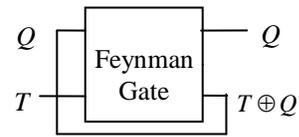

Fig. 2. Design of T Flip-Flop.

Our proposed T flip-flop with Q output has only one gate, quantum cost 1, delay 1 and no garbage outputs.

### B. Design of clocked T Flip-flop

The characteristic equation of a clocked T flip-flop is $Q = (T \oplus Q).CLK \oplus \overline{CLK}.Q$. The equation can be simplified as $Q = (T.CLK) \oplus Q$. This clocked flip-flop is realized by a Peres gate and a Feynman gate. Two designs are proposed here. Fig.3 shows the design of a clocked T flip-flop. Fig.3 (a) is used for synchronous counter and Fig.3 (b) is used for asynchronous counter.

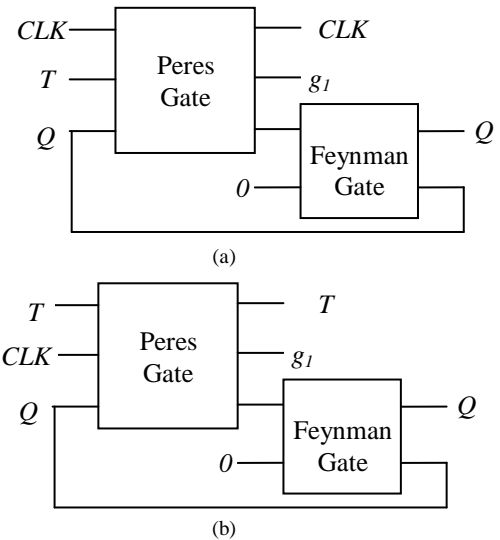

Fig. 3. Deign of clocked T flip-flop.

Both of our proposed designs have quantum cost 5, delay 5 and garbage output 1. Comparisons of the resources of clocked T flip-flop of our design with the existing design are given in Table I.

TABLE I. COMPARISONS OF DIFFERENT TYPES OF CLOCKED T FLIP-FLOPS

| Gated T flip-flop design | Cost Comparison | | |
|---|---|---|---|
| | *Quantum Cost* | *Delay* | *Garbage Outputs* |
| Proposed | 5 | 5 | 1 |
| Chuang[12] | 6 | 6 | 2 |
| Thapliyal[14] | 6 | 6 | 2 |

## C. Master Slave T flip-flop

To implement master slave T flip-flop, it needs one flip-flop working as master and another is slave. The same strategy is followed here. For master flip-flop, Peres gate is modified. The input vector $I_V$ and output vector $O_V$ of a 3*3 modified Peres gate, MPG are defined as follows, $I_v = (A, B, C)$ and $O_v = (P = \overline{A}, Q = A \oplus B, R = AB \oplus C)$. The quantum cost of MPG is 4. The block diagram and equivalent quantum representation for 3*3 MPG are shown in Fig.4.

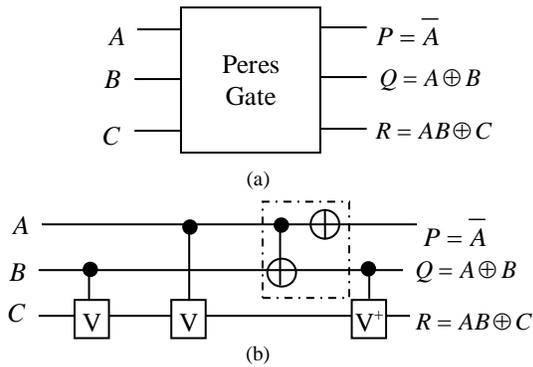

Fig. 4. (a) Block diagram of 3*3 MPG gate and (b) Equivalent quantum representation.

The modification is required to produce negative clocked pulse without generating any additional gate cost. The proposed master slave T flip-flop is shown in Fig.5.

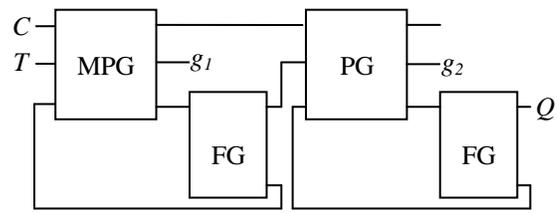

Fig. 5. Design of master slave T flip-flop.

Our proposed design has quantum cost 10, delay 10 and garbage output 2. Comparisons of resources of master slave T flip-flop of our design with the existing design are given in Table II.

TABLE II. COMPARISONS OF DIFFERENT TYPES OF MASTER SLAVE T FLIP-FLOPS

| Master-Slave T flip-flop | Cost Comparison | | |
|---|---|---|---|
| | *Quantum Cost* | *Delay* | *Garbage Outputs* |
| Proposed | 10 | 10 | 2 |
| Thapliyal[14] | 11 | 11 | 3 |
| Thapliyal [18] | 17 | 17 | 4 |

## D. Design of Asynchronous Counter

In asynchronous counter, the T flip-flops are arranged in such a way that output of one flip-flop is connected to the clock input of the next higher order flip-flop. The output of a flip-flop triggers the next flip-flop. The flip-flop holding the least significant bit receives the incoming count pulse. Our proposed 4 bit asynchronous counter is shown in Fig.6. The counter is realized by 4 Peres gates and some Feynman gates.

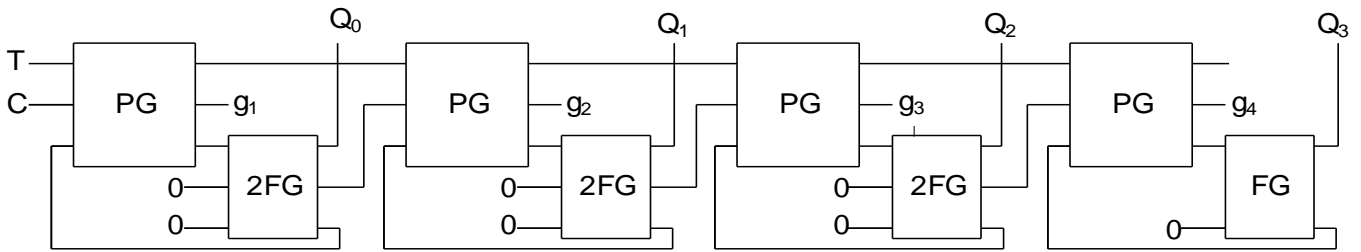

Fig. 6. Design of 4bit asynchronous counter

Our proposed 4bit asynchronous counter has quantum cost 23, delay 23 and garbage output 4. There is not much good result available about asynchronous counter. Comparisons of our proposed design with existing result are shown in Table III.

TABLE III. COMPARISONS OF PROPOSED DESIGN OF 4 BIT ASYNCHRONOUS COUNTER WITH EXISTING DESIGN

| 4bit asynchronous counter | Cost Comparison | | |
|---|---|---|---|
| | *Quantum Cost* | *Delay* | *Garbage Outputs* |
| Proposed | 23 | 23 | 4 |
| Rajmohan[17] | 55 | 55 | 12 |

**Theorem 1:** To construct n bit asynchronous counter, if g is the total number of gates required to design the counter producing b number of garbage outputs then g≥2n and b≥n.

**Proof:** Each flip-flop consists of two gates; n bit counter requires n number of flip-flops. No additional gates are required to interconnect each other. So total number of gates required to design the counter is 2n, hence g≥2n. Similarly, every flip-flop produces only one garbage output. No garbage output produced while interconnection among flip-flops. So the total number of garbage out is n, hence b≥n.

**Theorem 2:** The quantum cost of an n bit asynchronous counter is $Q_n \geq 6n-1$.

**Proof:** For n=1, only one Peres gate and one Feynman gate is required to construct the counter. The quantum cost of Peres gate is 4 and quantum cost of Feynman gate is 1. So the total cost 4+1=5.

Now for n>1, one Peres gate and one double Feynman gate is required for each flip-flop in the counter except the last one which requires one Feynman gate instead of double Feynman gate. So for n bit asynchronous counter it needs n Peres gate, (n-1) double Feynman gate and one Feynman gate. The quantum cost of double Feynman gate is 2. So the total quantum cost is 4*n+2(n-1)+1 =6n-1, Hence $Q_n \geq 6n-1$.

*E. Design of Synchronous Counter*

Synchronous counter is different from asynchronous counter in that clock pulses are applied to the inputs of all the flip-flops at a time. A flip-flop is complemented depending on the input value T and the clock pulse. The flip-flop in least significant position is completed with every clock pulse. A flip-flop in other position is complemented only when all the outputs of preceding flip-flops produces 1. Same strategy is followed here to implement the synchronous counter. Fig.7 shows our proposed 4bit synchronous counter.

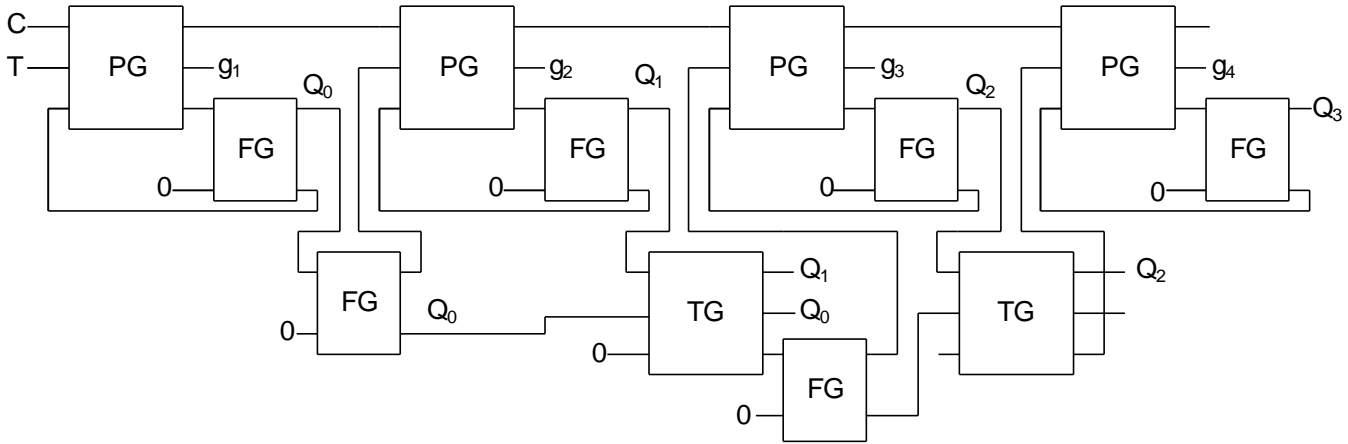

Fig. 7. Design of 4bit synchronous counter

Our proposed 4bit synchronous counter has quantum cost 32, delay 32 and garbage output 4. . Comparisons of our proposed design with existing result are shown in Table IV.

TABLE IV. COMPARISONS OF PROPOSED DESIGN OF 4 BIT SYNCHRONOUS COUNTER WITH EXISTING DESIGN

| 4bit synchronous counter | Cost Comparison | | |
|---|---|---|---|
| | *Quantum Cost* | *Delay* | *Garbage Outputs* |
| Proposed | 32 | 32 | 4 |
| khan[16] | 35 | 35 | 4 |

**Theorem 3:** To construct n ($\geq$3) bit synchronous counter, if g is the total number of gates required to design the counter producing b number of garbage outputs then g$\geq$4n-4 and b$\geq$n.

**Proof:** Each flip-flop consists of two gates; n bit counter requires n number of flip-flops. For n=3, one Toffoli and one Feynman is required to carry out all the outputs to the next higher positioned flip-flop. So total number of gates required is 3*2+2=8.

For n>3, 2n number of gates required for the flip-flops and 2(n-2) number of gates are required to carry out all the lower outputs to the next higher outputs. So the total number of gates required is 2n+2(n-2)=4n-4. Every flip-flop produces only one garbage output. No garbage output produced while interconnection among flip-flops and to carry out outputs to next higher flip-flop. So the total number of garbage out is n, hence b$\geq$n.

**Theorem 4:** The quantum cost of an n($\geq$3) bit synchronous counter is $Q_n \geq 11n-12$.

**Proof:** For n ($\geq$3) bit synchronous counter it requires n flip-flops. Each flip-flop consists of one Peres gate and one Feynman gate. Additional (n-2) Toffoli gates and (n-2) Feynman gates are required to carry out all the outputs to the next higher flip-flop. So it requires n number of Peres gate, (n-2) number of Toffoli gates and n+(n-2) =2n-2 number of Feynman gates. Quantum cost of Peres gate is 4, Quantum cost of Toffoli gate is 5 and quantum cost of Feynman gate is 1. So total quantum cost = 4*n + 5*(n-2)+1*(2n-2)=11n-12, hence $Q_n \geq 11n-12$.

V. CONCLUSION

Reducing quantum cost in sequential circuit is always a challenging one. Only a few attempts were made on reversible counter. A novel reversible design for both n bit synchronous and asynchronous counter is proposed. Appropriate algorithms and theorems are presented to clarify the proposed design and to establish its efficiency. As compared to the best reported designs in literature, the proposed designs are better in terms of quantum cost, delay and garbage outputs. The proposed design can have great impact in reversible computing.

ACKNOWLEDGMENT

REFERENCES

[1] Rolf Landauer, "Irreversibility and Heat Generation in the Computing Process", IBM Journal of Research and Development, vol. 5, pp. 183-191, 1961.
[2] Charles H.Bennett, "Logical Reversibility of computation", IBM Journal of Research and Development, vol. 17, no. 6, pp. 525-532, 1973.
[3] Perkowski, M., A.Al-Rabadi, P. Kerntopf, A. Buller, M. Chrzanowska-Jeske, A. Mishchenko, M. Azad Khan, A. Coppola, S. Yanushkevich, V. Shmerko and L. Jozwiak, "A general decomposition for reversible logic", Proc. RM'2001, Starkville, pp: 119-138, 2001

[4] J.E Rice, "A New Look at Reversible Memory Elements", Proceedings International Symposium on Circuits and Systems(ISCAS) 2006, Kos, Greece, May 21-24 ,2006, pp. 243-246.
[5] Dmitri Maslov and D. Michael Miller, "Comparison of the cost metrics for reversible and quantum logic synthesis", http://arxiv.org/abs/quant-ph/0511008, 2006
[6] Mohammadi,M. and Mshghi,M, On figures ofmerit in reversible and quantumlogic designs, Quantum Inform. Process. 8, 4, 297–318, 2009.
[7] Md. SelimAl Mamun and Syed Monowar Hossain. "Design of Reversible Random Access Memory." *International Journal of Computer Applications* 56.15 (2012): 18-23.
[8] Richard P.Feynman, "Quantum mechanical computers," Foundations of Physics, vol. 16, no. 6, pp. 507-531, 1986.
[9] A. Peres, "Reversible Logic and Quantum Computers," Physical Review A, vol. 32, pp. 3266-3276, 1985.
[10] Md. Selim Al Mamun and David Menville, Quantum Cost Optimization for Reversible Sequential Circuit, (IJACSA) International Journal of Advanced Computer Science and Applications, Vol. 4, No. 12, 2013.
[11] Tommaso Toffoli, "Reversible Computing," Automata, Languages and Programming, 7th Colloquium of Lecture Notes in Computer Science, vol. 85, pp. 632-644, 1980.
[12] M.-L. Chuang and C.-Y. Wang, "Synthesis of reversible sequential elements," ACM journal of Engineering Technologies in Computing Systems (JETC). Vol. 3, No.4, 1–19, 2008.
[13] J. E. Rice, An introduction to reversible latches. The Computer journal,Vol. 51, No.6, 700–709. 2008.
[14] Himanshu Thapliyal and Nagarajan Ranganathan, Design of Reversible Sequential Circuits Optimizing Quantum Cost, Delay, and Garbage Outputs, ACMJournal onEmerging Technologies inComputer Systems,Vol. 6,No. 4,Article 14, Pub. date:December 2010.
[15] Siva Kumar Sastry, Hari Shyam Shroff, Sk.Noor Mahammad, V. Kamakoti", Efficient Building Blocks for Reversible Sequential Circuit Design" 1-4244-0173-9106/$20.00©2006IEEE
[16] Mozammel H A Khan and Marek Perkowski, Synthesis of Reversible Synchronous Counters, 2011 41st IEEE International Symposium on Multiple-Valued Logic, 0195-623X/11 $26.00 © 2011 IEEE
[17] V.Rajmohan, V.Ranganathan,"Design of counter using reversible logic" 978-1-4244-8679-3/11/$26.00 ©2011 IEEE.
[18] H. Thapliyal and A. P. Vinod, "Design of reversible sequential elements with feasibility of transistor implementation" In Proc. the 2007 IEEE Intl. Symp. On Cir.and Sys., pages 625–628, New Orleans, USA, May 2007.